\documentclass[aoas,MSNbibl,nameyear,secthm,seceqn,dvips]{arximspdf}
\usepackage{graphicx}

\doi{10.1214/11-AOAS480}
\volume{5}
\issue{2B}
\pubyear{2011}
\firstpage{1132}
\lastpage{1158}

\makeatletter
\let\epsilon\varepsilon

  \let\sv@tabnotetext\tabnotetext
  \let\sv@tabnotemark@fmt\tabnotemark@fmt
   \long\def\legend#1{{\let\tabnote@indent\leavevmode\sv@tabnotetext[]{}{#1}}}

\def\@bmisc[#1]{%
  \get@battribute{unstr}%
  \common@pub@types%
  \let\bauthor\bbl@bauthor%
  \let\bhowpublished\@firstofone%
  \def\borganization##1{{\bauthor@style ##1}}%
}

\makeatother

\begin{document}
\begin{frontmatter}

\title{Two-stage empirical likelihood for longitudinal neuroimaging data}
\runtitle{Empirical likelihood for longitudinal neuroimaging data}

\begin{aug}
\author[A]{\fnms{Xiaoyan} \snm{Shi}},
\author[A]{\fnms{Joseph G.} \snm{Ibrahim}\thanksref{aut1}},
\author[B]{\fnms{Jeffrey} \snm{Lieberman}},
\author[A]{\fnms{Martin}~\snm{Styner}\thanksref{aut2}},
\author[C]{\fnms{Yimei} \snm{Li}}
and
\author[A]{\fnms{Hongtu} \snm{Zhu}\corref{}\ead[label=e1]{hzhu@bios.unc.edu}\thanksref{aut1}}
\runauthor{X. Shi et al.}
\affiliation{University of North Carolina at Chapel Hill, University of North Carolina at Chapel Hill, Columbia
University, University of North Carolina at\\ Chapel Hill, St. Jude Children's Research Hospital\\ and University of North Carolina at Chapel Hill}
\thankstext{aut1}{Supported in part by
NIH Grants RR025747, CA142538, MH086633, AG033387, GM70335, and
CA74015.}
\thankstext{aut2}{Supported in part by UNC Neurodevelopmental Disorders Research Center HD 03110,
NIH NIBIB grant P01 EB002779, Eli Lilly user initiated information
technology Grant PCG
TR:033107, and the NIH Roadmap Grant U54 EB005149-01, National Alliance
for Medical
Image Computing.}
\address[A]{X. Shi\\
J. G. Ibrahim\\
M. Styner\\
H. Zhu\\
Department of Biostatistics and\\
\quad Biomedical Research Imaging Center\\
University of North Carolina at Chapel Hill\\
Chapel Hill, North Carolina 27599-7420\\
USA\\
\printead{e1}} 
\address[B]{J. Lieberman\\
New York State Psychiatric Institute\\
1051 Riverside Drive\\
New York, New York 10032\\
USA}
\address[C]{Y. Li\\
St. Jude Children's Research Hospital\\
262 Danny Thomas Place\\
Memphis, Tennessee 38105-3678\\
USA}
\end{aug}

\received{\smonth{11} \syear{2009}}
\revised{\smonth{3} \syear{2011}}

%
\begin{abstract}
Longitudinal imaging studies are essential to
understanding the neural development of neuropsychiatric disorders,
substance use disorders, and the normal brain.
The main objective of this paper is to
develop a two-stage adjusted
exponentially tilted empirical likelihood (TETEL) for the spatial
analysis of neuroimaging data from longitudinal
studies. The TETEL method as a frequentist approach allows us to
efficiently analyze longitudinal data without
modeling temporal correlation and to classify different time-dependent
covariate types.
To account for spatial dependence, the TETEL method developed here
specifically combines all the data in the closest neighborhood of each voxel
(or pixel) on a 3-dimensional (3D) volume (or 2D surface) with
appropriate weights to calculate adaptive parameter estimates and
adaptive test statistics. Simulation studies are used to examine
the finite sample performance of the adjusted exponential tilted
likelihood ratio statistic and TETEL.
We demonstrate the application of our statistical methods to the
detection of the difference in the morphological changes of the
hippocampus across time between schizophrenia patients and healthy
subjects in a longitudinal schizophrenia study.
\end{abstract}

%

\end{frontmatter}

\section{Introduction}

Neuroimaging data, including both anatomical and\break functional magnetic
resonance imaging (MRI), have been/are being widely collected to understand
the neural development of neuropsychiatric disorders, substance use
disorders, and the normal brain
in various
longitudinal studies [Almli et al. (\citeyear{AlmRiv})]. For instance,
various morphometrical measures of the
morphology of the cortical and subcortical structures (e.g.,
hippocampus) are extracted from anatomical MRIs for understanding
neuroanatomical differences in brain structure across different
populations and across time.
Studies of brain morphology have been conducted widely to
characterize differences in brain structure across groups of healthy
individuals and persons with various diseases, and across time
[\citet{ThoTog02}, \citet{ThoCanTog02},
Styner et al. (\citeyear{Styetal05}), Zhu et al. (\citeyear{Zhuetal08N1})]. Moreover,
functional MRI (fMRI) is a valuable tool for understanding
functional integration of different brain regions in response to
specific stimuli and behavioral tasks and detecting the association
between brain function and covariates of interest, such as
diagnosis, behavioral task, severity of disease, age, or IQ
[\citet{Fri07}, Rogers et al. (\citeyear{Rogetal07}), \citet{HueSonMcC}].

Much effort has been devoted to developing frequentist and Bayesian
methods for analyzing neuroimaging data using numerical simulations and
theoretical
reasoning. Frequentist statistical methods for analyzing
neuroimaging data are often sequentially executed in two
steps. The first step involves fitting a general linear model
or a linear mixed model to neuroimaging data
from all subjects at each voxel
[Beckmann, Jenkinson and Smith (\citeyear{BecJenSmi03}),
Friston et al. (\citeyear{Frietal05}), \citet{Row05},
Woolrich et al. (\citeyear{Wooetal04}), Zhu et al. (\citeyear
{Zhuetal08N1})]. The
second step is to calculate adjusted $p$-values that account for
testing the hypotheses across multiple brain regions or across many
voxels of the imaging volume using various statistical methods
(e.g., random field theory, false discovery rate, or permutation
method) [\citet{CaoWor01}, Friston et al. (\citeyear{Frietal96}),
Hayasaka et al. (\citeyear{Hayetal04}),
\citet{LogRow04}, Worsley et al. (\citeyear{Woretal04})].
Most of these frequentist methods have been implemented in existing
neuroimaging software platforms,
including statistical parametric mapping (SPM)
(\href{http://www.fil.ion.ucl.ac.uk/spm/}{www.fil.ion.ucl.ac.uk/spm/}) and FMRIB Software Library (FSL)
(\href{http://www.fmrib.ox.ac.uk/fsl/}{www.fmrib.ox.ac.uk/fsl/}),
among many others.
In the recent literature, a~number of papers have been published on the development of Bayesian
spatial--temporal models for
functional imaging data [\citet{PenFlaTru07},
Bowman et al. (\citeyear{Bowetal08}), Woolrich et al.
(\citeyear{Wooetal04}), \citet{LuoPut05}]. Most Bayesian
approaches, however, are less practical due to
the extensively computational burden of running a Markov chain Monte
Carlo method in a large number of voxels [Bowman et al. (\citeyear
{Bowetal08})], and,
thus, they are limited to small or moderate anatomic regions and a
small number of regions of interest (ROI). Moreover, as pointed out
in \citet{SnoPleBea07}, the major drawbacks of ROI analysis include
the instability of statistical results obtained from ROI analysis and
the partial volume effect in relative large ROIs.

Existing statistical methods in the neuroimaging literature have two
major limitations for
analyzing longitudinal neuroimaging data, as
explained below. The respective strategies to resolve these two
limitations are detailed in Section \ref{sec2}.
The first limitation is that the parametric models, such as linear
mixed models, require the correct specification of the temporal
correlation structure and
cannot properly distinguish between different types of time-dependent
covariates (types I, II and III)
[Diggle et al. (\citeyear{Digetal02}), \citet{LaiSma07},
\citet{PepAnd94}].
A distinctive feature of longitudinal neuroimaging data is that it is
able to
characterize individual change in neuroimaging measurements (e.g.,
volumetric and morphometric) over time, and the time-dependent
covariates of
interest may influence
change. Imaging measurements
of the same individual usually exhibit positive
correlation and the strength of the correlation decreases with the
time separation [Liang and Zeger (\citeyear{LiaZeg86})].
Moreover, longitudinal data may provide crucial information for a
causal role of a time-dependent covariate (e.g., exposure)
in the disease process [Diggle et al. (\citeyear{Digetal02}),
\citet{LaiSma07}, \citet{PepAnd94}].
Improperly handling time-dependent covariates and ignoring (or
incorrectly modeling) temporal correlation structure in imaging
measures likely would influence subsequent statistical inference,
such as increasing the false positive and negative
errors, and result in misleading scientific inferences [Diggle et al.
(\citeyear{Digetal02}), \citet{LaiSma07}].

The second limitation is that most smoothing methods apply the same
amount of smoothing throughout the whole image,
which can be problematic near the
edges of the significant regions.
Although it is common to apply a~smoothing step before applying a
voxel-wise
approach for the analysis of neuroimaging
data [\citet{PolMaz94}, Shafie et al. (\citeyear{Shaetal03}),
\citet{LinWag08}],
the voxel-wise
method suffers from the same amount of smoothing throughout the whole
image and
the arbitrary
choice of smoothing extent [Hecke et al. (\citeyear{Hecetal09}),
Jones et al. (\citeyear{Jonetal05})]. Jones et al.
(\citeyear{Jonetal05}) have shown that the final results of voxel-based
analysis can strongly
depend on the amount of smoothing in the smoothed diffusion imaging
data. Recently, \citet{YueLohLin10}
introduced a spatially smoothing method using nonstationary spatial
Gaussian Markov random fields to spatially
and adaptively smooth images. Their approach, however,
can be computationally extensive for 3D imaging data.

In this paper we will develop
strategies to resolve these two limitations.
To resolve the first limitation, we develop an adjusted exponentially
tilted empirical likelihood method, called AETEL, for the analysis of
longitudinal neuroimaging data with time-dependent
covariates. AETEL is a nonparametric method that is built on a set of
estimating equations and the number of estimating equations can be
larger than the number of parameters.
Thus, it avoids parametric assumptions and this feature is very
appealing for the analysis of real
neuroimaging data, such as brain morphological measures, because
the distribution of the
univariate (or multivariate) neuroimaging measurements often
deviates from the Gaussian distribution [\citet{AshFri00},
Salmond et al. (\citeyear{Saletal02}), \citet{LuoNic03}].
Using more estimating equations than the number of parameters allows us to
appropriately handle time-dependent covariates of different types and to
make an efficient
use of the estimating equations without the need of
modeling the temporal correlation in longitudinal data [\citet
{LaiSma07}, \citet{QuLinLi00}].
AETEL also provides a natural test statistic to test
whether a specific covariate is of a certain type (types I, II and III).

To resolve the second limitation,
we develop a two-stage
AETEL, abbreviated as
TETEL, for
the analysis of longitudinal neuroimaging data.
TETEL integrates a
smoothing method into our AETEL for
carrying out statistical inference on neuroimaging data.
The TETEL method, as an adaptive procedure,
fits AETEL at each voxel in stage 1. Then, TETEL uses the
information learned from stage 1 to discard the data from the
neighboring voxels with dissimilar signal pattern
and to incorporate
the data from the neighboring voxels with similar signal pattern
to adaptively calculate parameter estimates and test statistics.
TETEL avoids
using the same amount of smoothing throughout the whole image in most
smoothing methods.
In addition, theoretically, we can establish consistency and asymptotic
normality of
the estimators and test statistics obtained from TETEL.

Section \ref{sec2} of this paper introduces the shape data of the hippocampus
structure from a longitudinal schizophrenia study and presents the new
statistical methods just
described. In Section \ref{sec3} we conduct simulation studies to examine
the finite sample performance of the TETEL
method. Section~\ref{sec4} illustrates an application of the proposed methods
to the longitudinal schizophrenia study of the hippocampus. We present
concluding remarks in Section \ref{sec5}.

\section{Data and methods}\label{sec2}

\subsection{Longitudinal schizophrenia study of hippocampus shape}

This is a~longitudinal, randomized, controlled, multisite,
double-blind study conducted at 14
academic medical centers in North America and western Europe, with
partial funding from Lilly Research Laboratories [Lieberman et al.
(\citeyear{Lieetal05}), Styner et al. (\citeyear{Styetal04})]. In this
study $238$ first-episode schizophrenia
patients were enrolled meeting
the following criteria: age 16 to 40 years; onset of psychiatric
symptoms before age 35; diagnosis of schizophrenia, schizophreniform,
or schizoaffective disorder according to the fourth edition of
diagnostic and statistical manual of mental disorders (DSM-IV)
criteria; and various
treatment and substance dependence conditions.
After random allocation at baseline, $123$ patients were selected to
receive a conventional antipsychotic, haloperidol (2--20 mg$/$d), and
$115$ were selected to receive an atypical antipsychotic, olanzapine
(5--20 mg$/$d). Patients were treated and followed up to 47 months. Also,
$56$ healthy control
subjects matched to the patient's demographic characteristics were enrolled.
Neurocognitive and MRI assessments were performed at months 0
(baseline), 3, 6, 13, 24, 36, and 47 approximately,
with different subjects having different visiting times, and some subjects
dropped out during the course of the study.

\begin{figure}

\includegraphics{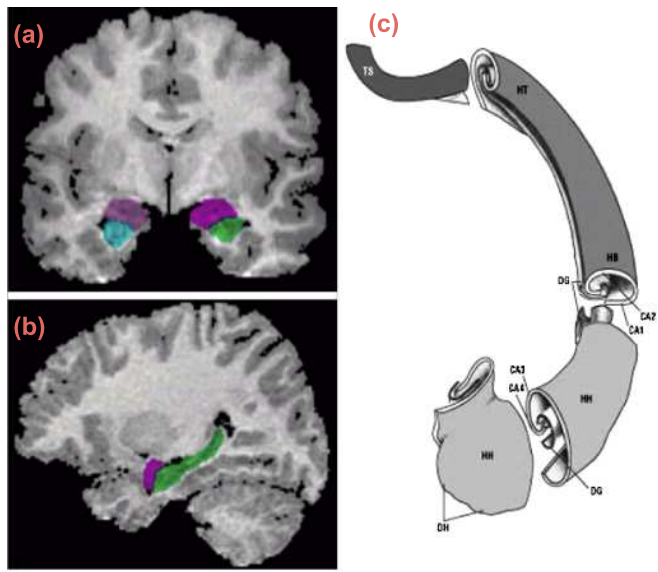}

\caption{Location of hippocampus in the context of the surrounding
structures in the coronal \textup{(a)} and sagittal \textup{(b)} views.
Subregions of the hippocampus in \textup{(c)} showing the head of the
hippocampus (HH), the digitationes hippocampi (DH), the hippocampal
body (HB), the hippocampal tail (HT), the terminal segment of the HT
(TS), the dentate gyrus (DG), and the fields of the cornu ammonis
(CA1--CA4). Adapted with permission from Springer Verlag, Heidelberg,
Germany [Duvernoy (\citeyear{Duv05})].}\label{fig1}
\end{figure}

The hippocampus, a gray matter structure
in the limbic system, is involved in processes of motivation and
emotions and has a central role in the formation of memory.
The hippocampus is a paired structure with mirror-image halves
in the left and right brain hemispheres and located inside the medial
temporal lobe (Figure \ref{fig1}).
Many MRI studies have reported the reduction of hippocampal volume
demonstrated in schizophrenia subjects and at onset of the first
episode of psychotic symptoms before effects associated with treatment
and disease chronicity
[Lieberman et al. (\citeyear{Lieetal05})].


%

The aim of this study
is to use the boundary and medial shape of the hippocampus to
examine whether hippocampal abnormalities are present in schizophrenia patients.
Statistical shape modeling and analysis have emerged as important
tools for understanding cortical and subcortical structures from
medical images [\citet{DryMar98}]. We consider two approaches for shape
representation including a
spherical harmonic description sampled into a triangulated surfaces
(SPHARM-PDM) and a medial shape description [Pizer et al. (\citeyear
{Pizetal03}),
\citet{StyGer03}].
The SPHARM-PDM can only represent objects of spherical topology,
whereas the medial representation provides information on a rich set
of features, including local thickness. These shape features are not
accessible by conventional volume-based morphometry and offer us a
great opportunity to address the weaknesses of conventional volumetric methods.

We consider two sets of responses of interest.
The first set of responses was based on the SPHARM-PDM representation
of hippocampal surfaces.
We use the SPHARM-PDM [Styner et al. (\citeyear{Styetal04})] shape
representation to
establish surface correspondence and align the surface location
vectors across all subjects. The sampled SPHARM-PDM is a smooth,
accurate, fine-scale shape representation (Figure~\ref{fig3}). The
hippocampal surfaces
of different subjects are thus represented by the same number of
location vectors (with each location vector consisting of the
spatial $x, y,$ and $z$ coordinates of the corresponding vertex on the
SPHARM-PDM surface) and are used as the second set of responses.
Covariates of interest are race
(Caucasian, African American, and others), age (in years), gender,
group (the schizophrenia group and the healthy control group) and time
(visiting times in months).

The second
set of responses was
the hippocampus m-rep thickness at the 24 medial atoms of the left and
the right brain (Figure \ref{fig4}).
The m-rep is a~linked set of medial
primitives named medial atoms, which are formed from two equal
length vectors and are composed of a position, a~radius, a~frame
implying the tangent plane to the medial manifold, and an object
angle [Styner et al. (\citeyear{Styetal04})]. The m-rep thickness is
the radius of each
medial atom.
Covariates of interest were WBV,
race (Caucasian, African American, and others), age (in years), gender,
diagnostic status (patient or control), and visiting times (in weeks).
This WBV measure includes gray and white matter, ventricular
cerebrospinal fluid, cisterns, fissures, and cortical sulci. The WBV is
commonly used as a covariate in statistical analyses to control for
scaling effects [Arndt et al. (\citeyear{Arnetal91})]. Particularly,
WBV is a~time-dependent covariate and may vary with the hippocampus
thickness measurement.

\subsection{Estimating equations for longitudinal data}

We consider a longitudinal study of imaging data with $n$ subjects,
where a $q\times1$ covariate $\mathbf{x}_{i,j}$
(e.g., age, gender, height, and brain volume) is obtained for the
$i$th subject at the $j$th time point $t_{ij}$ for $i=1, \ldots, n$
and $j=1, \ldots, m_i$. Thus, there are at least $\sum_{i=1}^n
m_i=N$ images in the study. Based on each image, we observe
or compute neuroimaging measures, denoted
by $\mathbf{Y}_{i}=\{\mathbf{y}_{ij}(d)\dvtx d\in\mathcal{D}, j=1,
\ldots,
m_i\}$, across all $m_i$ time points from the $i$th subject, where
$d$ represents a voxel (or atom, or point) on $\mathcal{D}$, a specific
brain region.
The imaging measure $\mathbf{y}_{ij}(d)$ at each voxel $d$ can be
either univariate
or multivariate. For example, the m-rep thickness is a
univariate measure, whereas the location vector of SPHARM-PDM is a
three-dimensional MRI measure at each point [\citet{StyGer03},
\citet{ChuDalDav07}]. For notational simplicity,
we assume that the $\mathbf{y}_{ij}(d)$ are univariate measures.

We temporarily drop voxel $d$ from our notation. At a specific
voxel $d$ in the brain region, $\mathbf{z}_i=\{(\mathbf{y}_{ij}, \mathbf
{x}_{ij})\dvtx j=1, \ldots, m_i\}$ is independent and
satisfies a moment condition
%
\begin{equation} \label{ETELIMG1}
E\{g(\mathbf{z}_i, \theta)\}={0}\qquad  \mbox{for } i=1, \ldots, n,
\end{equation}
where $\theta$ is a $p\times1$ vector, $g(\cdot, \cdot)$ is an
$r\times1$ vector of known functions with $r\geq p$, and $E$ denotes
the expectation with respect to the true distribution of all the
$\mathbf{z}_i$'s. Equation (\ref{ETELIMG1}) is often referred to as a
set of
unbiased estimating equations or moments model [\citet{QinLaw94},
\citet{Han82}].
The moments model (\ref{ETELIMG1}) is more general than most parametric
models including linear mixed model used for the analysis of neuroimaging
data [Worsley et al.
(\citeyear{Woretal04}), \citet{QinLaw94}, \citet{Han82},
\citet{Sch07}, \citet{Owe01}].

For longitudinal data, although the measurements from different
subjects are independent, those within the same subject may be
highly correlated.
The generalized estimating equations (GEE) assume a working
covariance matrix for $\mathbf{y}_i=(\mathbf{y}_{i1}, \ldots, \mathbf
{y}_{im_i})^T$
given by
$
V_i.
$
Let $E(\mathbf{y}_i)=\mu_i(\beta)=(\mu_{i1}(\beta), \ldots, \mu
_{im_i}(\beta))^T$ and $D_i(\beta)=\partial\mu_i(\beta)/\partial\beta$.
Under the assumption that $E\{D_i(\beta)^TV_i^{-1}[\mathbf{y}_i-\mu
_i(\beta)]\}=0$,
\citet{LiaZeg86} proposed to use an estimator, denoted by $\hat
\beta_{\mathrm{gee}}$, which
solves
a set of GEEs as follows:
%
\begin{equation}\label{hongtugee}
G(\beta)= \sum_{i=1}^n D_i(\beta)^TV_i^{-1}[\mathbf{y}_i-\mu_i(\beta
)]=\mathbf{0}.
\end{equation}

For longitudinal data with time-dependent covariates,
whether $E[g(\mathbf{z}_i,\break \theta)]=E\{D_i(\beta)^TV_i^{-1}[\mathbf
{y}_i-\mu_i(\beta)]\}$ equals zero or
not depends on the type of time-dependent covariates and the structure
of $V_i$ [\citet{LaiSma07}].
The time-dependent covariate $\mathbf{x}_{ij}$ is of type
I if
%
\begin{equation}\label{type1}
E\{\partial_\beta\mu_{is}(\beta)[\mathbf{y}_{ij}-\mu_{ij}(\beta)]\}=0\qquad
\mbox{for all } s, j=1, \ldots, m_i,
\end{equation}
where $\partial_\beta=\partial/\partial\beta$.
A sufficient condition for type I covariates is $E[y_{ij}|\mathbf
{x}_{ij}]=E[y_{ij}|\mathbf{x}_{i1}, \ldots, \mathbf{x}_{im_i}]$. For
type I
covariates, we can set $g(\mathbf{z}_i,
\theta)\!=\!D_i(\beta)^TV_i^{-1}[\mathbf{y}_i\!-\mu_i(\beta)]$ and show that
$E[g(\mathbf{z}_i, \theta)]=0$. If $V_i$ is the covariance matrix
of $\mathbf{y}_i$, then the estimator $\hat\beta_{\mathrm{gee}}$ is an
efficient estimator. However,
$\hat\beta_{\mathrm{gee}}$ is inefficient under a misspecified $V_i$.
To increase the efficiency, we may choose several
candidate working covariance matrices $M_{i}^{(1)}, \ldots,
M_{i}^{(s_0)}$ and
assume $V_i^{-1}=\sum_{k=1}^{s_0} \alpha_k M_{i}^{(k)}$ for some
unknown constants $\alpha_k$ [\citet{QuLinLi00}]. Then, following
\citet{QuLinLi00}, we consider a set of estimating equations given by
%
\begin{equation}
g(\mathbf{z}_i, \theta)= \pmatrix{
D_i(\beta)^TM_{i}^{(1)}[\mathbf{y}_i-\mu_i(\beta)]\cr
\vdots\cr
D_i(\beta)^TM_{i}^{(s_0)}[\mathbf{y}_i-\mu_i(\beta)]
}\qquad  \mbox{for } i=1, \ldots, n.
\end{equation}
In this case, the number of functions in $g(\mathbf{z}_i, \theta)$ is
$s_0q>q$, when $s_0>1$.

The time-dependent covariate $\mathbf{x}_{ij}$ is of type
II if
%
\begin{equation}\label{type2}
E\{\partial_\beta\mu_{is}(\beta)[\mathbf{y}_{ij}-\mu_{ij}(\beta)]\}=0
\qquad \mbox{for all } s\geq j, j=1,\ldots, m_i.
\end{equation}
A sufficient condition for type II covariates is
%
\begin{equation}\label{type2suf}
p(\mathbf{x}_{i, t+1}, \ldots, \mathbf{x}_{im_i}|\mathbf{y}_{it},
\mathbf{x}_{it})=p(\mathbf{x}_{i, t+1}, \ldots, \mathbf
{x}_{im_i}|\mathbf{x}_{it}).
\end{equation}
For type II covariates, we can set $g(\mathbf{z}_i,
\theta)=D_i(\beta)^T[\mathbf{y}_i-\mu_i(\beta)]$, in which an
independent working covariance matrix is used.
However, the estimator $\hat\beta_{\mathrm{gee}}$ based on
the independent working correlation matrix is
inefficient, since we do not use the information
contained in $E\{\partial_\beta\mu_{is}(\beta)[\mathbf{y}_{ij}-\mu
_{ij}(\beta)]\}=0$ for all $s>j$. To increase the
efficiency of the estimate, we choose a set of lower triangular
$m_i\times m_i$
matrices $L_{i}^{(1)}, \ldots, L_{i}^{(s_0)}$, and then we consider
estimating equations given by
%
\begin{equation}\label{type2L}
g(\mathbf{z}_i, \theta)= \pmatrix{
D_i(\beta)^TL_{i}^{(1)}[\mathbf{y}_i-\mu_i(\beta)]\cr
\vdots\cr
D_i(\beta)^TL_{i}^{(s_0)}[\mathbf{y}_i-\mu_i(\beta)]
}\qquad  \mbox{for } i=1, \ldots, n.
\end{equation}
In this case, the number of functions in $g(\mathbf{z}_i, \theta)$ is
$s_0q>q$, when $s_0>1$. Supposing that $m_1=\cdots=m_n$, we can set
$s_0=m_1(m_1+1)/2$ and $L_{i}^{(b)}=\mathbf{e}_{s}\mathbf{e}_j^T$ for
$s\geq j$ and $b=1, \ldots, s_0$, where
$\mathbf{e}_s$ is a $q\times1$ vector with the $s$th component 1 and 0
otherwise. Thus, similar to \citet{LaiSma07}, we are able to
pick $\partial_\beta\mu_{is}(\beta)[\mathbf{y}_{ij}-\mu_{ij}(\beta)]$
for all $s\geq j$.

The time-dependent covariate $\mathbf{x}_{ij}$ is of type
III if
%
\begin{equation}\label{type3}
E\{\partial_\beta\mu_{is}(\beta)[\mathbf{y}_{ij}-\mu_{ij}(\beta)]\}\not
=0\qquad  \mbox{for some } s>j.
\end{equation}
For type III covariates, we need to choose $V_i$ as a diagonal
matrix. For instance, if $V_i=\mathbf{I}_i$, where $\mathbf{I}_i$ is an
$m_i\times m_i$ identity matrix, then $g(\mathbf{z}_i,
\theta)=D_i(\beta)^T[\mathbf{y}_i-\mu_i(\beta)]$. Furthermore, if we
assume the specific form for the variances of all $\mathbf{y}_{ij}$,
then we may set $V_i=\operatorname{diag}(\operatorname{Cov}(\mathbf{y}_i))$.

An overall strategy to analyze models with time-dependent covariates
is first to assume that the time-dependent covariates are of type III.
Then we test whether the time-dependent covariates are of type II, and if
the test is not rejected, we can go on to test if they are of type I.
Once the type of all the time-dependent covariates is decided, we use
the corresponding estimating equations.
See Section \ref{sec4} for more details.

\subsection{Adjusted exponentially tilted empirical likelihood}

We consider a nonparametric method, called an exponentially tilted
empirical likelihood, to carry
out statistical inference about $\theta$ based on a set of estimating
equations $\{g(\mathbf{z}_i, \theta)\dvtx i=1, \ldots, n\}$ [\citet{Sch07}].
The exponentially tilted empirical likelihood (ETEL) method is a
combination of the exponentially tilted (ET) method and
the empirical likelihood (EL) method.
Both EL [\citet{Owe01}, \citet{QinLaw94}] and ET [\citet
{ImbSpaJoh98}] methods combine the
reliability of nonparametric methods with the effectiveness of the
likelihood approach. The EL estimator exhibits desirable higher-order
asymptotic properties, whereas
the EL estimator may fail to be $\sqrt{n}$-convergent in the presence
of model misspcification.
In contrast, the ETEL estimator maintains $\sqrt{n}$-convergence under
model misspecification
[\citet{Sch07}].

However, most empirical likelihood type methods including ETEL suffer
from two pitfalls: relatively
low precision of the chi-square approximation and
nonexistence of solutions to the estimating equations [\citet
{CheVarAbr08}, \citet{LiuChe10}].
\citet{CheVarAbr08} introduce a novel adjustment to these
empirical likelihood methods and develop an
iterative algorithm that converges very fast. Simulation studies have
shown that the adjusted empirical
likelihood methods perform as well as the linear regression model with
Gaussian noise when data
are symmetrically distributed, while the adjusted empirical likelihood
methods are superior when data
have skewed distribution [Zhu et al. (\citeyear{Zhuetal09}),
\citet{CheVarAbr08}, \citet{LiuChe10}].

Following \citet{CheVarAbr08}, we consider an adjustment of ETEL,
abbreviated as AETEL, by introducing an adjustment
%
\begin{equation}
{g}_{n+1}(\theta)=-\frac{a_n}n\sum_{i=1}^n g(\mathbf{z}_i, \theta),
\end{equation}
where $a_n=\max(1, \log(n)/2)$.
Then,
AETEL is defined as
%
\begin{equation}\label{hongtu}
\ell_{\mathrm{Aetel}}(\theta)=-(n+1)^{-1}\sum_{i=1}^{n+1}\log
\bigl((n+1)\hat p_i(\theta)\bigr),
\end{equation}
where $\hat p_i(\theta)$ is the solution to
\[
\min_{p_1, \ldots, p_{n+1}}(n+1)^{-1} \sum_{i=1}^{n+1} [(n+1) p_i]
\log[(n+1)p_i]
\]
subject to
\[
\sum_{i=1}^{n+1} p_i=1, p_i\geq0\quad  \mbox{and} \quad \sum_{i=1}^{n}
p_i g(\mathbf{z}_i, \theta)+p_{n+1} g_{n+1}(\theta)=0.
\]
The maximum AETEL estimator, denoted by
$\hat\theta_{\mathrm{Aetel}}$, minimizes a criterion given by
%
\begin{equation} \label{AETELeq2}
\hat\theta_{\mathrm{Aetel}}=\operatorname{argmin}\limits_{\theta}
\ell_{\mathrm{Aetel}}(\theta).
\end{equation}
According to a
duality theorem in convex analysis [\citet{NewSmi04}],
\[
\hat
p_{n+1}(\theta)= \frac{{\exp(\hat{t}(\theta)^T
g_{n+1}(\theta))}}{T_g(\theta)} \quad \mbox{and}\quad
\hat
p_{i}(\theta)=\frac{\exp(\hat{t}(\theta)^T g(\mathbf{z}_i,
\theta))}{T_g(\theta)}
\]
for $i=1, \ldots, n,$
in which
\begin{eqnarray*}
T_g(\theta)&=&\sum_{j=1}^{n} \exp(\hat{t}(\theta)^T
g(\mathbf{z}_j, \theta))+\exp(\hat{t}(\theta)^T g_{n+1}(\theta)),
\\
\hat{t}(\theta)&=&\operatorname{argmax}\limits_{t} \Biggl\{-\sum_{i=1}^n
\exp(-{t}^Tg(\mathbf{z}_i,
\theta))-\exp(-{t}^Tg_{n+1}(\theta))\Biggr\}.
\end{eqnarray*}
We use the numerical algorithm proposed by \citet{CheVarAbr08} to
compute $\hat\theta_{\mathrm{Aetel}}$,
which combines the modified Newton--Raphson algorithm and the simplex
method. Compared with that of computing ETEL, this numerical
algorithm of \citet{CheVarAbr08} converges very fast and the
solution to AETEL is guaranteed.

We consider testing the linear hypotheses:
%
\begin{equation}
H_0\dvtx R\theta=\mathbf{b}_0 \quad \mbox{vs.}\quad  H_1\dvtx R\theta\not= \mathbf{b}_0,
\end{equation}
where $ {R}$ is a $c_0\times p$ matrix of full row rank
and $ \mathbf{b}_0$ is a $c_0\times1$ specified vector.
Most scientific questions in neuroimaging studies can be formulated into
linear hypotheses, such as a comparison of brain regions across
diagnostic groups
and a detection of changes in brain regions across time.
The AETEL
ratio statistic for testing $ {R}\theta= {\mathbf{b}_0}$ can be
constructed as follows:
%
\begin{equation} \label{AETELeq4}
\mathit{LR}_{\mathrm{Aetel}}=-2(n+1)\Bigl\{\sup_{\theta\dvtx{R}\theta
={\mathbf{b}_0}}
\ell_{\mathrm{Aetel}}(\theta)- \sup_{\theta} \ell_{\mathrm
{Aetel}}(\theta)\Bigr\}.
\end{equation}
Thus, to compute $\mathit{LR}_{\mathrm{Aetel}}$, we also need to
compute the maximum
AETEL estimator, denoted by $\hat\theta_{\mathrm{Aetel}, 0}$,
subject to an additional constraint $R\theta=\mathbf{b}_0$.

Under some conditions on $g(\mathbf{z}_i, \theta)$,
we have the following theorem, whose detailed proof can be found in a
supplementary document [\citet{Shietal2011}].

\begin{thm}\label{thm1}
If assumptions \textup{(A1)--(A4)} in the supplementary document are true, then
we have the following:
\begin{longlist}[(a)]
\item[(a)] $\sqrt{n}(\hat\theta_{\mathrm{Aetel}}-\theta_0)$ converges to $\nu
_0=N(0, \Sigma)$
in distribution, where $\theta_0$ denotes the true value of $\theta$
and $\Sigma=(DV^{-1}D^T)^{-1}$, in which
\[
D=\lim_{n\rightarrow\infty}
n^{-1}\sum_{i=1}^n\partial_\theta g(\mathbf{z}_i, \theta) \quad \mbox{and}\quad
V=\lim_{n\rightarrow\infty}n^{-1}\sum_{i=1}^ng(\mathbf{z}_i,
\theta)^{\otimes2};
\]

\item[(b)] under the null hypothesis $H_0$, $\mathit{LR}_{\mathrm{Aetel}}$
converges to a
$\chi^2(c_0)$ distribution;

\item[(c)] if $E[g(\mathbf{z}_i, \theta)]\!=\!0$ for all $i$ and $r\!>\!p$, then
$\mathit{LR}_{\mathrm{GF}}\!=\!-2(n\!+\!1)\sup_{\theta} \ell_{\mathrm{Aetel}}(\theta)$
is asymptotically $\chi^2(r\!-\!p)$.
\end{longlist}
\end{thm}

We have established consistency and
asymptotic normality of $\hat\theta_{\mathrm{Aetel}}$ and the
asymptotic $\chi^2$ distribution of $\mathit{LR}_{\mathrm{Aetel}}$.
Theorem \ref{thm1} also shows that
AETEL has the same first-order asymptotic properties as
ETEL [\citet{Sch07}]. High-order precision of AETEL can be
explored by following the arguments in
\citet{LiuChe10}. It will be shown that the
chi-square approximation of the \mbox{AETEL} likelihood ratio
statistics is found precise, compared with the existing
ETEL [\citet{Owe01}, \citet{LiuChe10},
\citet{CheVarAbr08}]. Providing a reliable $p$-value
at each voxel is crucial for controlling the family-wise error rate and
false discovery rate (FDR) across the entire brain region [\citet
{BenHoc95}, Worsley et al. (\citeyear{Woretal04})].

\subsection{Two-stage adaptive estimation procedure}

We propose a two-stage adaptive estimation procedure for computing
parameter estimates and likelihood ratio statistics
for the spatial and adaptive analysis of neuroimaging data in 3D
volumes (or 2D surfaces). To distinguish data and parameter in
different voxels,
we introduce voxel $d$ into our notation. For instance, $\mathbf
{z}_i(d)$ and $\theta(d)$, respectively, denote the $i$th observation
and the parameter at voxel~$d$.

Stage 1 is to calculate the AETEL estimator of the
parameter ${{\theta}}(d)$, denoted by
$\hat{{\theta}}_{\mathrm{Aetel}}(d)$, based on a set of estimating
equations $\{g(\mathbf{z}_i(d),
{{\theta}}(d))\dvtx i=1, \ldots, n\}$ at each voxel $d\in{\mathcal
D}$.

One chooses a set of estimating equations $\{g(\mathbf{z}_i(d),
{{\theta}}(d))\dvtx i=1, \ldots, n\}$ according to a specific type of
time-dependent covariate
and then substitutes them into
(\ref{hongtu}) to build $\ell_{\mathrm{Aetel}}(\theta(d); d)$.
Subsequently, we
solve $\hat{{\theta}}_{\mathrm{Aetel}}(d)$ according to
(\ref{AETELeq2}) by minimizing $\ell_{\mathrm{Aetel}}(\theta(d); d)$,
and then we obtain\vspace*{2pt}
a set of parameter estimates $\{ \hat{\theta}_{\mathrm{Aetel}}(d)\dvtx
d\in{\mathcal D}\}$.

Stage 2 is to calculate the TETEL estimator of
${{\theta}}(d)$, denoted by $\hat{{\theta}}_{\mathrm{Tetel}}(d)$, by
utilizing the information contained in $\{ \hat{\theta}_{\mathrm
{Aetel}}(d)\dvtx d\in{\mathcal D}\}$. Then, we calculate a
TETEL ratio statistic, denoted by $\mathit{LR}_{\mathrm{Tetel}}(d)$,
for testing
$H_0(d)\dvtx\break R{{\theta}}(d)=\mathbf{b}_0$.

Specifically, one combines
all data in the voxel $d$ and the set of the closest neighboring voxels of
$d$, denoted by $N(d)$, to form a new set of estimating equations
$\{\tilde g(\mathbf{z}_i(d), {{\theta}}(d); d)\dvtx i=1, \ldots,
n\}$ as follows:
%
\begin{equation} \label{HongtuRev1}
\tilde g(\mathbf{z}_i(d), {{\theta}}(d); d)=
\sum_{d'\in N(d)\cup\{d\}} \omega(d'; d)g(\mathbf{z}_i(d'), {{\theta}}(d)),
\end{equation}
where $\omega(d'; d)$ is a weight
describing the similarity between voxel $d$ and any $d'\in N(d)$.
The weights $\omega(d'; d)$ at each $d$
depend on the parameters $\{\hat{{\theta}}_{\mathrm{Aetel}}(d')\dvtx
d'\in N(d)\cup\{d\}\}$ calculated in Stage
1. From now on, we assume that $\omega(d'; d)$ takes the form
%
\begin{equation} \label{similarity}
\omega(d'; d)=\exp(-\mathit{LR}_{\mathrm{Aetel}}(d'; d)/C_n),
\end{equation}
where $C_n=\chi_{1-\alpha}^2(p) \log(n)/5$ and
$\chi_{1-\alpha}^2(p)$ is the upper $\alpha$-percentile of the
$\chi^2(p)$ distribution. In addition,
%
\begin{equation}
\qquad \mathit{LR}_{\mathrm{Aetel}}(d'; d)=-2(n+1)\Bigl\{
\ell_{\mathrm{Aetel}}(\hat{\theta}_{\mathrm{Aetel}}(d'); d)- \sup
_{{\theta}} \ell_{\mathrm{Aetel}}({\theta}; d)\Bigr\},
\end{equation}
in which $\ell_{\mathrm{Aetel}}({{\theta}}; d)$ is defined in (\ref
{hongtu}) based on
the estimating equations $\{ g(\mathbf{z}_i(d), {{\theta}}(d))\dvtx
i=1, \ldots,
n\}$.
Statistically, $\mathit{LR}_{\mathrm{Aetel}}(d'; d)$
denotes the AETEL
ratio statistic for testing the hypothesis $H_0\dvtx{\theta}(d)= \hat
{\theta}_{\mathrm{Aetel}}(d')$. Note that
$\mathit{LR}_{\mathrm{Aetel}}(d';
d)\geq0$ and $\mathit{LR}_{\mathrm{Aetel}}(d; d)=0$, which yields
$\omega(d; d)=1$. If $\hat{\theta}_{\mathrm{Aetel}}(d')$ is close to
$\hat{\theta}_{\mathrm{Aetel}}(d)$, then $\mathit{LR}_{\mathrm
{Aetel}}(d'; d)$ is close to zero
and $\omega(d'; d)$ will be close to 1. However, if the difference
between $\hat{\theta}_{\mathrm{Aetel}}(d')$ and $\hat{\theta}_{\mathrm
{Aetel}}(d)$ is large,
then $\mathit{LR}_{\mathrm{Aetel}}(d'; d)$ is large and $\omega(d'; d)$
will be small.
Thus, $\omega(d'; d)$ defined in (\ref{similarity}) truly characterizes
the similarity between voxels $d$ and $d'$.

One substitutes $ \tilde g(\mathbf{z}_i(d), {{\theta}}(d); d)$ in (\ref
{HongtuRev1}) into
(\ref{hongtu}) to build a new function, denoted by $\ell_{\mathrm
{Tetel}}(\theta(d); d)$, and then
solves $\hat{{\theta}}_{\mathrm{Tetel}}(d)$ according to
(\ref{AETELeq2}) by minimizing $\ell_{\mathrm{Tetel}}(\theta(d); d)$.
Finally, to test
$H_0(d)\dvtx R{{\theta}}(d)=\mathbf{b}_0$, one uses $ \tilde g(\mathbf
{z}_i(d), {{\theta}}(d); d)$ in (\ref{HongtuRev1}) to calculate the TETEL
ratio statistic $\mathit{LR}_{\mathrm{Tetel}}(d)$ according to (\ref
{AETELeq4}). Note that the key difference between $\mathit{LR}_{\mathrm
{Tetel}}(d)$ and $\mathit{LR}_{\mathrm{Aetel}}(d)$
lies in their different sets of estimating equations.

Although the two-stage procedure only combines the data in the voxels of
$N(d)$ with the data in voxel $d$,
they may
preserve the long-range correlation structure in the imaging
data, because
the
neighborhoods of all voxels are
consecutively connected. Thus, the two-stage procedure captures a
substantial amount of spatial information in the imaging data.
For the sake of space, we only present the asymptotic properties of
$\hat{{\theta}}_{\mathrm{Tetel}}(d)$ and $\mathit{LR}_{\mathrm
{Tetel}}(d)$ below.

\begin{thm}\label{thm2}
If assumptions \textup{(A1)--(A3)} and \textup{(A5)--(A7)} in the
supplementary document are true, then we have the following:
\begin{longlist}[(a)]
\item[(a)] $\sqrt{n}(\hat{{\theta}}_{\mathrm{Tetel}}(d)-{{\theta}}_0(d))$
converges to $\nu(d)=N(0, \Sigma(d))$ in distribution, where
${{\theta}}_0(d)$ is the true value of ${{\theta}}(d)$ in the
voxel $d$ and $\Sigma(d)=[D(d)V(d)^{-1}\times D(d)^T]^{-1}$, in which
\[
D(d)=\lim_{n\rightarrow\infty}
n^{-1}\sum_{i=1}^n\partial_{\theta} \tilde g(\mathbf{z}_i(d), {{\theta
}}_0(d); d)
\]
and
\[
V(d)=\lim_{n\rightarrow\infty}n^{-1}\sum_{i=1}^n\tilde g(\mathbf
{z}_i(d), {{\theta}}_0(d); d)^{\otimes2};
\]

\item[(b)] under the null hypothesis $H_0(d)$, $\mathit{LR}_{\mathrm
{Tetel}}(d)$ converges
in distribution to a $\chi^2(c_0)$ distribution.
\end{longlist}
\end{thm}

Theorem \ref{thm2} establishes the asymptotic consistency and normality of
$\hat{\theta}_{\mathrm{Tetel}}(d)$ and the asymptotic $\chi^2$
distribution of $\mathit{LR}_{\mathrm{Tetel}}(d)$. Theorem \ref{thm2}
also shows that the
asymptotic variance of $\hat{\theta}_{\mathrm{Tetel}}(d)$ depends on all
the data in $N(d)\cup\{d\}$ for all subjects. Since the weights
$\omega(d'; d)$ automatically put large weights on the neighboring
voxels with similar pattern and
small weights on the neighboring voxels with dissimilar pattern, it
follows that the TETEL procedure
produces more accurate parameter estimates and more powerful test
statistics.

TETEL has three features. TETEL not only downweights
the data from the neighboring voxels with dissimilar signal pattern,
but also
incorporates the data from the neighboring voxels with similar signal
pattern to
adaptively calculate parameter estimates and test statistics. TETEL avoids
using the same amount of smoothness throughout the whole image in most
smoothing methods.
Our theoretical results ensure the asymptotic consistency and
normality of $\hat\theta_{\mathrm{Tetel}}(d)$
and the asymptotic $\chi^2$ distribution of\break $\mathit{LR}_{\mathrm{Tetel}}(d)$.
Then, we can approximate the $p$-value of $\mathit{LR}_{\mathrm
{Tetel}}(d)$ at each voxel~$d$.
Finally, we correct for multiple comparisons by using either the
family-wise error rate or
false discovery rate (FDR) across the entire brain region [\citet
{BenHoc95}, Worsley et al. (\citeyear{Woretal04})]. Since the smoothing
stage in TETEL
usually introduces the positive dependency\break among all $\mathit
{LR}_{\mathrm{Tetel}}(d)$, it allows us to apply FDR in \citet
{BenYek01} to
control the false discovery
rate.

\section{Simulation studies}\label{sec3}
$\!\!$Three sets of simulation studies were conducted to examine the finite sample
performance of our AETEL and
TETEL methods.

\subsection{Study I: Longitudinal data}

We considered the following model:
%
\begin{equation}\label{long}
\mathbf{y}_{ij}=\beta_0+\beta_1 t_{ij}+\beta_2 x_{i}+\beta_3
t_{ij}x_{i}+b_i+\epsilon_{ij}
\end{equation}
for $i=1,\ldots,n$, where $t_{ij}$ denotes time taking values in
$(1,2,3,4,5)$, $x_{i}$ was independently generated from a $N(0,1)$, and
$b_{i}$ was independently generated from a $N(0,1)$.
Errors ${\epsilon}_{ij}$ were
independently generated from $N( 0, 1)$ and $\chi^2(3)-3$, respectively,
where $\chi^2(3)$ represents a chi-squared random variable with three
degrees of freedom. The $\chi^2(3)-3$ distribution is very skewed and
differs substantially from any symmetric distribution, such as a
Gaussian distribution.
The true value of $(\beta_{0},\beta_{1},\beta_{2})^T$
was set at $(1,1,1)^T$ and $\beta_{3}$ was varied as 0, 0.05, 0.10,
0.15, and 0.20.
We tested the hypothesis $H_0\dvtx\beta_3=0$ vs. $H_1\dvtx
\beta_3\not=0$ using $\mathit{LR}_{\mathrm{Aetel}}$.
To assess both Type I and II error rates of $\mathit{LR}_{\mathrm{Aetel}}$,
we used generalized estimating
equations assuming an exchangeable working
correlation matrix to construct $\mathit{LR}_{\mathrm{Aetel}}$ and
then compared it with the ETEL likelihood
ratio statistic, denoted by $\mathit{LR}_{\mathrm{Etel}}$, and
the Wald statistic, denoted by $W_n$, obtained from
the ``true'' linear mixed model (\ref{long}) representing an ideal scenario.
We considered $n=40,$ $60$, and $80$.
The 1,000 replications were used to calculate the estimates of
rejection rates with significance level $\alpha=5\%$.

The type I error rates of $\mathit{LR}_{\mathrm{Aetel}}$ and $W_n$ are
reasonably accurate for all sample sizes ($n =
40, 60,$ or $80$) considered and for all different distributions of
error terms at the 5$\%$ significant level (Table 1).
In contrast, the type I error rates of $\mathit{LR}_{\mathrm{Etel}}$ are
slightly inflated for $n=40$.
The type II error rates for $\mathit{LR}_{\mathrm{Aetel}}$ and $W_n$
are similar under both error distributions and for all sample sizes
(Table 1).
However, the power of the three test statistics to reject the null
hypothesis increases modestly when the distribution of the error
terms follows the skewed distribution $\chi^2(3)-3$ (Table \ref{table1}). This
decline in the type II error rate was caused by the fact that the
variance of $\chi^2(3)-3$ is larger than that of $N(0, 1)$.
Compared with $\mathit{LR}_{\mathrm{Aetel}}$ and $W_n$, $\mathit
{LR}_{\mathrm{Etel}}$ has slightly larger power, which may be due to its
inflated type I error rates.
Consistent with our expectation, the statistical power for rejecting
the null hypothesis increases with the sample size $n$.

%
\begin{table}
\caption{Simulation study for comparing
$\mathit{LR}_{\mathrm{Aetel}}$, $\mathit{LR}_{\mathrm{Etel}}$, and $W_n$ for
testing $H_0\dvtx\beta_3=0$ against $H_1\dvtx
\beta_3\not=0$}\label{table1}
\begin{tabular*}{\tablewidth}{@{\extracolsep{\fill}}lccccccc@{}}
\hline
& & \multicolumn{3}{c}{$\bolds{\chi^2(3)-3}$} & \multicolumn{3}{c@{}}{$\bolds{N(0, 1)}$}\\[-5pt]
&& \multicolumn{3}{c}{\hrulefill}& \multicolumn{3}{c@{}}{\hrulefill}\\
& & $\bolds{n=40}$ & $\bolds{n=60}$& $\bolds{n=80}$ & $\bolds{n=40}$ & $\bolds{n=60}$ & $\bolds{n=80}$ \\
\hline
$\beta_{3}=0.0$ &$\mathit{LR}_{\mathrm{Etel}}$ & 0.078 & 0.066 & 0.059 & 0.082 &
0.070 & 0.058\\
&$\mathit{LR}_{\mathrm{Aetel}}$ & 0.066 & 0.054 & 0.055 & 0.068 & 0.064
& 0.058\\
&$W_n$& 0.062 & 0.064 & 0.068 & 0.078 & 0.064 & 0.054 \\[3pt]
$\beta_{3}=0.05$ &$\mathit{LR}_{\mathrm{Etel}}$ & 0.112 & 0.118 & 0.118 & 0.186
& 0.280 & 0.300\\
& $\mathit{LR}_{\mathrm{Aetel}}$ & 0.088 & 0.104 & 0.102 & 0.156 &
0.254 & 0.278\\
&$W_n$ & 0.094 & 0.102 & 0.100 & 0.164 & 0.244 & 0.264\\[3pt]
$\beta_{3}=0.10$ &$\mathit{LR}_{\mathrm{Etel}}$ & 0.198 & 0.182 & 0.286 & 0.548
& 0.866 & 0.804\\
&$\mathit{LR}_{\mathrm{Aetel}}$ & 0.176 & 0.160 & 0.268 & 0.506 & 0.848
& 0.792\\
&$W_n$ & 0.168 & 0.164 & 0.270 & 0.474 & 0.786 & 0.728\\[3pt]
$\beta_{3}=0.15$ & $\mathit{LR}_{\mathrm{Etel}}$ & 0.280 & 0.340 & 0.394 & 0.986
& 0.930 & 0.978\\
&$\mathit{LR}_{\mathrm{Aetel}}$ & 0.250 & 0.316 & 0.374 & 0.986 & 0.920
& 0.974\\
&$W_n$ & 0.268 & 0.324 & 0.356 & 0.978 & 0.892 & 0.970\\[3pt]
$\beta_{3}=0.20$ &$\mathit{LR}_{\mathrm{Etel}}$ & 0.560 & 0.520 & 0.720 & 0.996
& 1\hphantom{.000} & 1\hphantom{.000}\\
&$\mathit{LR}_{\mathrm{Aetel}}$ & 0.532 & 0.488 & 0.702 & 0.990 & 1\hphantom{.000} &
1\hphantom{.000}\\
&$W_n$& 0.512 & 0.482 & 0.682 & 0.982 & 0.998 & 1\hphantom{.000}\\
\hline
\end{tabular*}
\legend{Estimates of
rejection rates were reported for $N(0, 1)$ and $\chi^2(3)-3$
distributed data at 3 different sample sizes
($n=40, 60, 80$) at significance level $\alpha=5\%$. For each case,
1,000 simulated
data sets were used.}
\vspace*{3pt}
\end{table}

\subsection{Study II: Testing the type of time-dependent covariates}

We used the simulation study for a type II time-dependent covariate
in Section 4.1 of \citet{LaiSma07} to examine the finite sample
performance of
our AETEL method. The data
were simulated under the mechanism
\[
{y}_{it}=\beta_{0}+\beta_{1}x_{it}+\beta_{2}x_{i,t-1}+b_{i}+e_{it}
\quad \mbox{and}\quad  x_{it}=\beta_3 x_{i,t-1}+\epsilon_{it},
\]
where
$b_{i},e_{it}$, and $\epsilon_{it}$ are mutually independent and
normally distributed with mean 0 and variances $4$, $1$, and $1$,
respectively; the $x_{it}$-process is stationary, that is, $x_{i0}\sim
N(0, \sigma^2_\epsilon/(1-\beta_3^2))$. We refer the reader to
\citet{LaiSma07} for more details.
Note that $x_{it}$ is a type II covariate. We used our
AETEL method with the following estimating equations: (a) the type II
estimating equations according to (\ref{type2}), labeled type II;
(b) the type III estimating equations according to (\ref{type3}),
labeled type III; (c) GEE using the independent working
correlation, labeled GEE independence; (d) GEE using the
exchangeable working correlation, labeled GEE exchangeable; (e) GEE
using the autoregressive AR-1 working correlation, labeled GEE AR-1. We compared
the bias, root-mean-square error, and the efficiency of each case for
the parameter $\beta_1$ to the GEE independence case (the
efficiency is the ratio of the mean-square error of the GEE
independence case to that of the case).\looseness=-1

As we can see from Table
\ref{t:four}, GEE independence and GEE AR-1 are biased, because they
use some invalid estimating equations. The other three are all
unbiased, with type II being more efficient than the other two.
Combining all available valid estimating equations does improve
efficiency.
With the same type II estimating equations, our method has slightly
less RMSE ($0.0401$ vs. $0.0407$) than \citet{LaiSma07}'s
method.

%
\begin{table}[b]
\vspace*{3pt}
\caption{Results of AETEL with various estimating equations
for a type II time-dependent covariate}
\label{t:four}
\begin{tabular*}{\tablewidth}{@{\extracolsep{\fill}}lccc@{}}
\hline
\textbf{Estimating equations} & \textbf{Bias} & \textbf{RMSE} & \textbf{Efficiency}\\
\hline
Type II & $\hphantom{-}0.00$ & $0.040$ & $1.82$\\
Type III & $\hphantom{-}0.00$ & $0.053$ & $1.04$\\
GEE independence & $\hphantom{-}0.00$ & $0.054$ & $1.00$\\
GEE exchangeable &$-0.12$ & $0.104$ & --\\
GEE AR-1 &$-0.79$ & $0.661$ & --\\
\hline
\end{tabular*}
\end{table}


\subsection{Study III: Spatial data}

%
\begin{figure}

\includegraphics{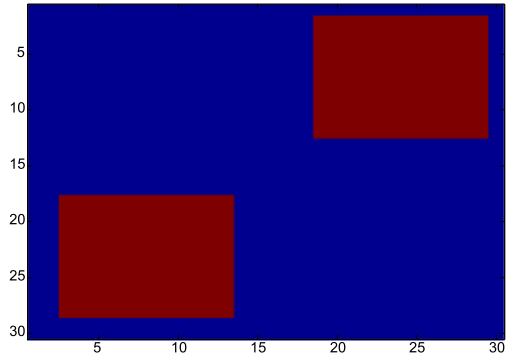}
\vspace*{-5pt}
\caption{Two red regions of interest (ROIs) on a $30\times30$ image.
The ROIs are indicated by the red area.}\label{fig2}
\vspace*{-5pt}
\end{figure}

We simulated data at all $m=900$ pixels on a $30\times30$ phantom image
(Figure \ref{fig2}). At a given voxel $d$,
%
\begin{equation}
\mathbf{y}_{ij}(d)=\beta_0(d)+\beta_1(d) t_{ij}+\beta_2(d) x_{i}+\beta
_3(d) t_{ij}x_{i}+b_i(d)+\epsilon_{ij}(d)
\end{equation}
for $i=1,\ldots,n$ and $j=1,\ldots,m_i$, where $t_{ij}$ is the time
taking values in $(1,2,3,4,5)$, $x_{i}$ was independently generated
from a $N(0,1)$, and $b_{i}(d)$ was independently generated from a
$N(0,1)$.
Errors ${\epsilon}_{ij}(d)$ were
independently generated from $N( 0, 1)$ and $\chi^2(3)-3$, respectively.
We tested the hypotheses $H_0\dvtx\beta_3(d)=0$ and $H_1\dvtx
\beta_3(d)\neq0$ across all pixels.
To assess the Type I and II error rates at the pixel level, we set
$\beta_0(d)=\beta_1(d)=\beta_2(d)=0$ across all pixels $d$ and
varied $\beta_3(d)$ as 0.0, 0.05, 0.10, 0.15, and 0.20.
Specifically, we created two
regions of interest (ROI) by setting $\beta_3(d)$ as 0.05, 0.10, 0.15,
and 0.20,
and setting $\beta_3(d)=0$ outside of the two ROIs
in order to assess the finite sample performance of our method at
different signal-to-noise ratios (SNRs).
We considered $n=40$ and $80$.

We used generalized
estimation equations with an exchangeable working correlation matrix to
calculate
$\hat\theta(d)$ and $\mathit{LR}_{\mathrm{Aetel}}(d)$ in Stage 1. In
Stage 2 we used the four first-order neighbors of pixel $d$ to form
$N(d)$ and then calculated $\mathit{LR}_{\mathrm{Tetel}}(d)$.
As a comparison with the conventional analysis on image data,
we first smoothed image data by using the heat kernel smoothing method
with 16 iterations, which gave
an effective smoothness of about 4~pixels [\citet{ChuDalDav07}],
and then calculated the Wald statistic based on
GEE with an exchangeable working correlation matrix at each pixel.
The
100 replications were used to approximate rejection rate with
significance level $\alpha=5\%$.

%
\begin{table}
\caption{Comparison of the two stages of TETEL for unsmoothed spatial
data and the Wald test statistic for smoothed spatial data:
true average rejection rates for voxels inside the ROI and false
average rejection rates for voxels outside of the ROI
were reported for $N(0, 1)$ and $\chi^2(3)-3$ distributed data, and 2
different sample sizes
($n=40$ and $80$) at $\alpha=5\%$. For each case, 100~simulated~data~sets were used}\label{t:spatial2}
\begin{tabular*}{\tablewidth}{@{\extracolsep{\fill}}lccccccccc@{}}
\hline
& &\multicolumn{4}{c}{$\bolds{\mathit{LR}_{\mathrm{Tetal}}}$} & \multicolumn
{4}{c@{}}{\textbf{Wald}} \\[-5pt]
&&\multicolumn{4}{c}{\hrulefill}&\multicolumn{4}{c@{}}{\hrulefill}\\
& &\multicolumn{2}{c}{$\bolds{n=40}$} & \multicolumn{2}{c}{$\bolds{n=80}$}& \multicolumn
{2}{c}{$\bolds{n=40}$} & \multicolumn{2}{c@{}}{$\bolds{n=80}$} \\[-5pt]
&&\multicolumn{2}{c}{\hrulefill}&\multicolumn{2}{c}{\hrulefill}&\multicolumn{2}{c}{\hrulefill}&\multicolumn{2}{c@{}}{\hrulefill}\\
$\bolds{\beta_3}$ & \textbf{Stage} &\textbf{True}&\textbf{False}&\textbf{True}&\textbf{False}& \textbf{True}&\textbf{False} &\textbf{True}&\textbf{False} \\
\hline
&&\multicolumn{8}{c@{}}{$N(0, 1)$} \\
0.05 & Stage 1 & 0.223 & 0.088 & 0.329 & 0.068 & 0.711 & 0.101 & 0.891
& 0.105 \\
& Stage 2 & 0.302 & 0.089 & 0.426 & 0.069 & & & & \\
0.10 & Stage 1 & 0.571 & 0.087 & 0.820 & 0.069 & 0.964 & 0.15 & 0.991 &
0.158 \\
& Stage 2 & 0.690 & 0.088 & 0.910 & 0.070 & & & & \\
0.15 & Stage 1 & 0.863 & 0.089 & 0.984 & 0.069 & 0.996 & 0.184 & 0.998
& 0.177 \\
& Stage 2 & 0.954 & 0.090 & 0.998 & 0.069 & & & & \\
0.20 & Stage 1 & 0.987 & 0.089 & 0.999 & 0.069 & 0.999 & 0.193 & 0.999
& 0.192 \\
& Stage 2 & 0.992 & 0.090 & 1.000 & 0.069 & & & & \\[3pt]
&&\multicolumn{8}{c@{}}{$\chi^2(3)-3$} \\
0.05 & Stage 1 & 0.117 & 0.085 & 0.122 & 0.070 & 0.313 & 0.089 & 0.331
& 0.073 \\
& Stage 2 & 0.212 & 0.090 & 0.232 & 0.070 & & & & \\
0.10 & Stage 1 & 0.193 & 0.087 & 0.259 & 0.069 & 0.567 & 0.099 & 0.858
& 0.099 \\
& Stage 2 & 0.278 & 0.089 & 0.411 & 0.070 & & & & \\
0.15 & Stage 1 & 0.313 & 0.090 & 0.447 & 0.068 & 0.847 & 0.113 & 0.948
& 0.123 \\
& Stage 2 & 0.486 & 0.091 & 0.649 & 0.070 & & & & \\
0.20 & Stage 1 & 0.463 & 0.090 & 0.660 & 0.069 & 0.947 & 0.130 & 0.979
& 0.145 \\
& Stage 2 & 0.653 & 0.090 & 0.859 & 0.069 & & & & \\
\hline
\end{tabular*}
\vspace*{-8pt}
\end{table}

As shown in Table \ref{t:spatial2}, the Type I rejection rates outside of ROIs for both
$\mathit{LR}_{\mathrm{Aetel}}$ and $\mathit{LR}_{\mathrm{Tetel}}$
are relatively accurate for all cases, while the statistical power for
rejecting the null hypothesis
in ROIs significantly increases with the absolute value of $\beta
_3(d)$. Compared with
$\mathit{LR}_{\mathrm{Aetel}}$, $\mathit{LR}_{\mathrm{Tetel}}$ has
higher statistical power for rejecting the null hypothesis in
ROIs with $\beta_3(d)\not=0$. In contrast, compared with
$\mathit{LR}_{\mathrm{Aetel}}$ and $\mathit{LR}_{\mathrm{Tetel}}$
based on the unsmoothed imaging data,
although the Wald statistic for the smoothed imaging data has higher
statistical power for rejecting the null hypothesis in ROIs,
its Type I error rate is inflated and increases with the absolute
value of $\beta_3(d)$.
The decline in the type I and II error rates is caused by the fact
that the variance of $\chi^2(3)-3$ is larger than that of $N(0, 1)$.
We also tried different degrees of smoothness and ROIs with different
sizes and found
that the degree of smoothness and ROI size can have profound effect on
the Type I and II error rates of the Wald statistic (not presented here).

\section{Hippocampus shape}\label{sec4}

\subsection{Hippocampus SPHARM-PDM representation}

Let $\mathbf{y}_{ij}(d)$ be the $3\times1$ coordinate vector at voxel
$d$ on the left and right hippocampus SPHARM-PDMs and $\mathbf
{x}_{ij}=(1, \mbox{gender}_{i}, \mbox{age}_{i}, \mbox{SC1}_{i}, \mbox{SC2}_{i},
\mathit{race}1_{i}, \mathit{race}2_{i}, \mathit{time}_{ij})^T$, where
$\mbox{SC1}$ and $\mbox{SC2}$ were, respectively, dummy
variables for haloperidol-treated SC patient and olanzapine-treated
SC patient versus healthy controls, and $\mathit{race}1$ and $\mathit{race}2$ were,
respectively, dummy variables for Caucasian and African American
versus other race.
Let $\mathbf{y}_i(d)=(\mathbf{y}_{i1}(d)^T, \ldots,
\mathbf{y}_{im_i}(d)^T)^T$ and $A \otimes B$ denote the Kronecker
product of matrices $A$ and $B$.
We assume that the mean and covariance matrix of $\mathbf{y}_i(d)$ are,
respectively, given by
\[
E(\mathbf{y}_{i}(d))=
\pmatrix{\mathbf{x}_{i1}^T\otimes\mathbf{I}_3\cr\cdots\cr
\mathbf{x}_{im_i}^T\otimes\mathbf{I}_3
}\beta(d)\!\!\!
\quad \mbox{and}\!\!\!\quad
\operatorname{Cov}(\mathbf{y}_{ij}(d))=V_i(d)=R_{i}(\alpha(d)) \otimes\Sigma(d),
\]
where $\beta(d)$ is a $24
\times1$ vector,
$R_{i}(\alpha(d))=(\alpha(d)^{|j-k|})$ is the standard autoregressive
AR-1 correlation matrix and
$\Sigma(d)$ is a $3\times3$ covariance matrix of $\mathbf{y}_{ij}(d)$.
We estimated $\alpha(d)$ and $\Sigma(d)$ by using Pearson residuals,
which were calculated by
solving GEEs with an independent working correlation matrix. For now
on, $V_i(d)$ [or $\alpha(d)$ and $\Sigma(d)$] are assumed to be known.
For the data analysis, we used the moment model based on GEE in (\ref
{hongtugee}) since there is no time-dependent covariate except time
itself. The $g(\mathbf{z}_i(d), \theta(d); d)$ which is used in TETEL
is given by
\[
g(\mathbf{z}_i(d), \theta(d); d)=\sum_{i=1}^{n}
\pmatrix{\mathbf{x}_{i1}^T\otimes\mathbf{I}_3\cr\cdots\cr
\mathbf{x}_{im_i}^T\otimes\mathbf{I}_3
}^TV_i(d)^{-1} \left[\mathbf{y}_i(d)-
\pmatrix{\mathbf{x}_{i1}^T\otimes\mathbf{I}_3\cr\cdots\cr
\mathbf{x}_{im_i}^T\otimes\mathbf{I}_3
}\beta(d)\right].
\]
Existing statistical methods
of image data in SPM require that the error
distribution is Gaussian and the variance is constant. The
Shapiro--Wilk test rejects the normality assumption at many voxels of
both the left and right hippocampus structures, and, thus, our
nonparametric TETEL method is preferred for the
analysis of this data set.

%
\begin{figure}[t!]

\includegraphics{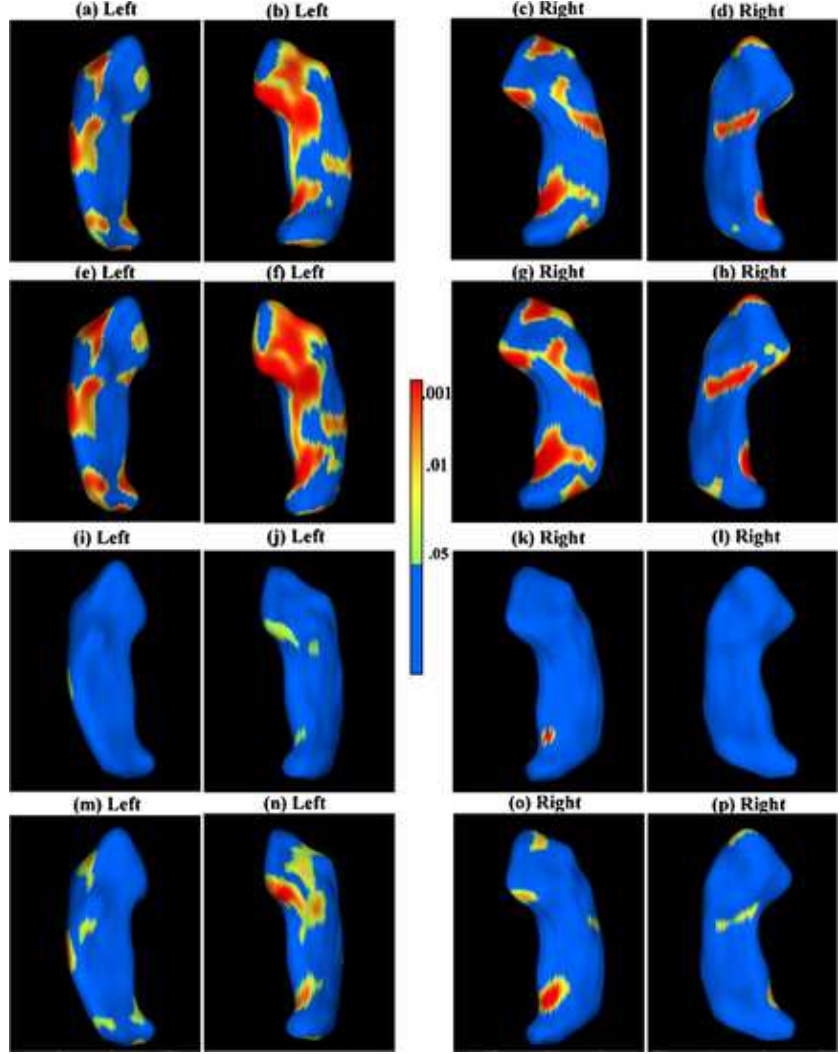}

\caption{Results from the longitudinal schizophrenia
study. The first and third rows are for the first stage
($\mathit{LR}_{\mathrm{Aetel}}$): the color-coded raw $p$-value maps of group
effect for the left hippocampus \textup{(a, b)} and the right
hippocampus \textup{(c,
d)}, and the corresponding color-coded corrected $p$-value maps of
group effect for the left hippocampus \textup{(i, j)} and the right hippocampus
\textup{(k, l)}. The second and fourth rows are for the second stage
($\mathit{LR}_{\mathrm{Tetel}}$): the color-coded $p$-value maps of
group effect for
the left hippocampus \textup{(e, f)} and the right hippocampus \textup
\mbox{(g, h)}, and
the corresponding color-coded corrected $p$-value maps of group effect
for the left hippocampus \textup{(m, n)}
and the right hippocampus \textup{(o, p)}.}\label{fig3}
\end{figure}

Since our goal is to detect the difference in the SPHARM-PDM surface
shape between the schizophrenia and control groups, we used
$\mathit{LR}_{\mathrm{Aetel}}$ and $\mathit{LR}_{\mathrm{Tetel}}$ to
carry out the test. Moreover, in Stage 2, we used
the closest neighbors of each voxel $d$ to form $N(d)$. The color-coded
$p$-values of the $\mathit{LR}_{\mathrm{Aetel}}$ and $\mathit
{LR}_{\mathrm{Tetel}}$ and their corrected
$p$-values using FDR across the voxels of both the left and right
reference hippocampi are shown in Figure 3 [\citet{BenYek01}], in
which the top row is
for the first stage ($\mathit{LR}_{\mathrm{Aetel}}$) and the bottom row
is for the
second stage ($\mathit{LR}_{\mathrm{Tetel}}$).

The analyses
show strong shape differences in the superior, anterior parts of the
left hippocampus, at the intersection of cornu ammonis 1 and cornu
ammonis~2, previously not
shown. Posterior shape changes at the hippocampal tail shown in
chronic schizophrenics [Styner et al. (\citeyear{Styetal04})] are
detected here already in
first episode patients. Furthermore, the results also confirm those
reported in Narr et al. (\citeyear{Naretal04}) by indicating a strong
medial shape
difference in the central, left hippocampal body in first episode
patients. Comparing the first and second rows, it is clear that
TETEL shows advantages in detecting more significant and smoother
activation areas.

\subsection{Hippocampus m-rep thickness}

\begin{figure}

\includegraphics{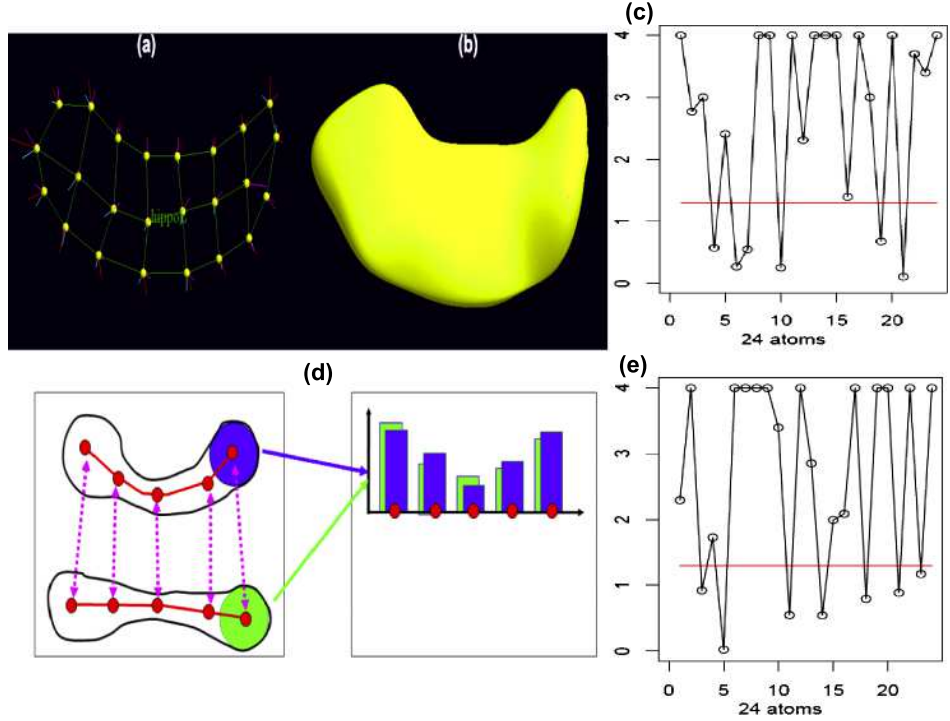}
\vspace*{-3pt}
\caption{An m-rep model of a hippocampus: \textup{(a)} an m-rep model
of the hippocampus; \textup{(b)}~the boundary
surface of the m-rep model of hippocampus; \textup{(d)} m-rep radius
(or thickness) measures at the
five atoms from two m-rep objects;
\textup{(c)} shows the
$-\log_{10}(p)$-values for the Shapiro--Wilk test for the residuals
at each atom on the left hippocampus; \textup{(e)} shows the
$-\log_{10}(p)$-values for the Shapiro--Wilk test for the residuals
at each atom on the right hippocampus. The red horizontal line is the
$0.05$ cutoff line.}\label{fig4}
\vspace*{-4pt}
\end{figure}

First, we considered the baseline analysis. We used the moment model
based on the estimating equations $\mathbf{x}_{i1} (\mathbf
{y}_{i1}-\mathbf{x}_{i1}^T\beta)$, where $\mathbf{y}_{i1}$ is the m-rep
thickness
measured at baseline for the $i$th subject at each medial atom
of the left and right hippocampi; $\mathbf{x}_{i1}$ is an $8\times1$
vector given by $\mathbf{x}_{i1}=(1, \mbox{gender}_{i}, \mbox{age}_{i},
\mbox{SC1}_{i}, \mbox{SC2}_{i}, \mathit{race}1_{i}, \mathit{race}2_{i}, WBV_{i1})^T$ and
$\beta=(\beta_0, \beta_1,\ldots,\beta_7)^T$.
Existing statistical methods of image data in SPM require that the
error distribution is Gaussian and the variance is constant.
The Shapiro--Wilk normality test was applied to check this parametric
assumption of the general linear model at each atom for the left
hippocampus and right hippocampus using the residuals. Figure \ref
{fig4}(c) and (e) show
that the Shapiro--Wilk test rejects the normality assumption at many
atoms of both the left and right hippocampus structures, therefore,
our nonparametric AETEL method is preferred for the analysis
of this data set.

%

%
\begin{figure}

\includegraphics{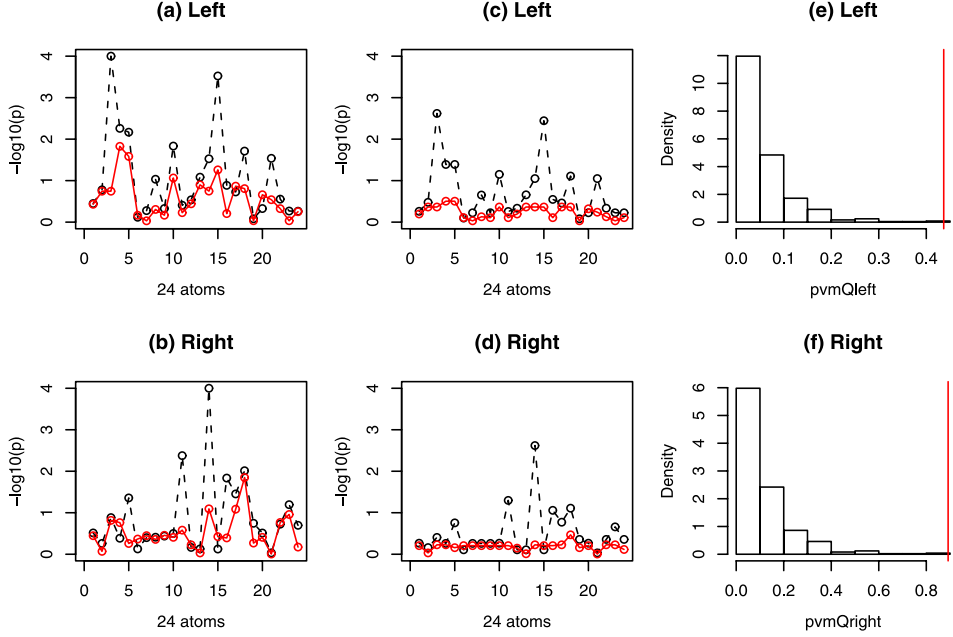}

\caption{An m-rep model of a hippocampus: Maps of
$-\log_{10}(p)$-values for testing WBV as
a type I time-dependent covariate (black) and a type II
time-dependent covariate (red): \textup{(a)} uncorrected
$-\log_{10}(p)$-values for left hippocampus; \textup{(b)} uncorrected
$-\log_{10}(p)$-values for right hippocampus; \textup{(c)} corrected
$-\log_{10}(p)$-values for left hippocampus; \textup{(d)} corrected
$-\log_{10}(p)$-values for right hippocampus; \textup{(e)} the
goodness-of-fit test
for the equation
$E\{\partial_\beta\mu_{i2}(\beta)[\mathbf{y}_{i3}-\mu_{i3}(\beta)]\}=0$
for the $3$rd atom on the left hippocampus; \textup{(f)} the goodness-of-fit
test for the equation
$E\{\partial_\beta\mu_{i2}(\beta)[\mathbf{y}_{i3}-\mu_{i3}(\beta)]\}=0$
for the $14$th atom on the right hippocampus.}\label{fig5}
\end{figure}

Here our goal is to detect differences in thickness of the
hippocampus across the three groups. Hence, we set the null hypotheses
$H_0\dvtx\beta_3=\beta_4=0$ at all 24 atoms for both the left and right
hippocampi. Accordingly, we have
\[
R= \pmatrix{
0&0&0&1&0&0&0&0\cr
0&0&0&0&1&0&0&0
}
\]
and $\mathbf{b}_0=(0,0)^T$. We used $\mathit{LR}_{\mathrm{Aetel}}$ to
carry out the test. The
color-coded $p$-values of the $\mathit{LR}_{\mathrm{Aetel}}$ across the
atoms of both the
left and right reference hippocampi are shown in Figure \ref{fig5}(a)
and (b).
The false discovery rate approach was used to correct for multiple
comparisons, and the resulting adjusted $p$-values were shown in Figure
\ref{fig5}(c) and (d). Before correcting for multiple comparisons,
there was
a significant group difference in m-rep thickness at the upper
central atoms in the left hippocampus and some area in the right
hippocampus. However, there is no significant group effect at any
atom after correcting for multiple comparisons.

Second, we did a longitudinal data analysis. The advantage of a
longitudinal study over
a baseline study is that it allows us to determine (i) whether the
change patterns of the
response are similar or not across the three groups; (ii) whether, on
average over time,
there is a difference in the response across the three groups. We
considered the moment model with $\mathbf{x}_{ij}=(1, \mbox{gender}_{i},
\mbox{age}_{i}, \mbox{SC1}_{i}, \mbox{SC2}_{i}, \mathit{race}1_{i},\mathit{race}2_{i},
\mathit{WBV}_{ij}, \mathit{time}_{ij}$, $\mbox{SC1}_{i}*\mathit{time}_{ij},\mbox{SC2}_{i}*\mathit{time}_{ij})^T$.

Since the WBV is a time-dependent covariate, we need to verify its
appropriate type. Moreover, from a neuroscience point of view, the
m-rep thickness at each atom serves as a local volumetric measure
and covaries with WBV. We started with type III and used
GEE in (\ref{hongtugee}) with $V_i=I_i$. Then we
used the type II equations specified in (\ref{type2}) and tested
whether WBV is type II against type III. The $\mathit{LR}_{\mathrm{GF}}$ did not
reject for almost all 24 atoms, suggesting WBV is a type II
covariate for most atoms. Furthermore, we used the type I equations
specified in (\ref{type1}) and tested whether WBV is type I against
type II. The $\mathit{LR}_{\mathrm{GF}}$ rejected that WBV was of type I for most
atoms (Figure \ref{fig5}). This indicates the invalidity of some type I
equations.
We used goodness-of-fit statistics in Zhu et al. (\citeyear
{Zhuetal08N2}) to test whether some of the extra equations
added for type I,
such as
\[
E\{\partial_{\beta_l}\mu_{is}(\beta)[\mathbf{y}_{ij}-\mu_{ij}(\beta)]\}
=0 \qquad \mbox{for all }
s<j, j=1, \ldots, m_i,
\]
were not valid. For instance, for the $3$rd atom on the left
hippocampus, the $p$-value of the goodness-of-fit test
for the newly added equation\break $E\{\partial_{\beta_l}\mu_{i2}(\beta
)[\mathbf{y}_{i3}-\mu_{i3}(\beta)]\}=0$ is smaller than $0.001$ [Figure
\ref{fig5}(e)]; for the $14$th atom on the right hippocampus, the
$p$-value of the goodness-of-fit test
for the newly added equation $E\{\partial_{\beta_l}\mu_{i2}(\beta
)[\mathbf{y}_{i3}-\mu_{i3}(\beta)]\}=0$ is
smaller than $0.001$ [Figure \ref{fig5}(f)]. Therefore, we treated WBV
as a type II time-dependent covariate and
used the corresponding estimating equations for the longitudinal data analysis.

%
\begin{figure}

\includegraphics{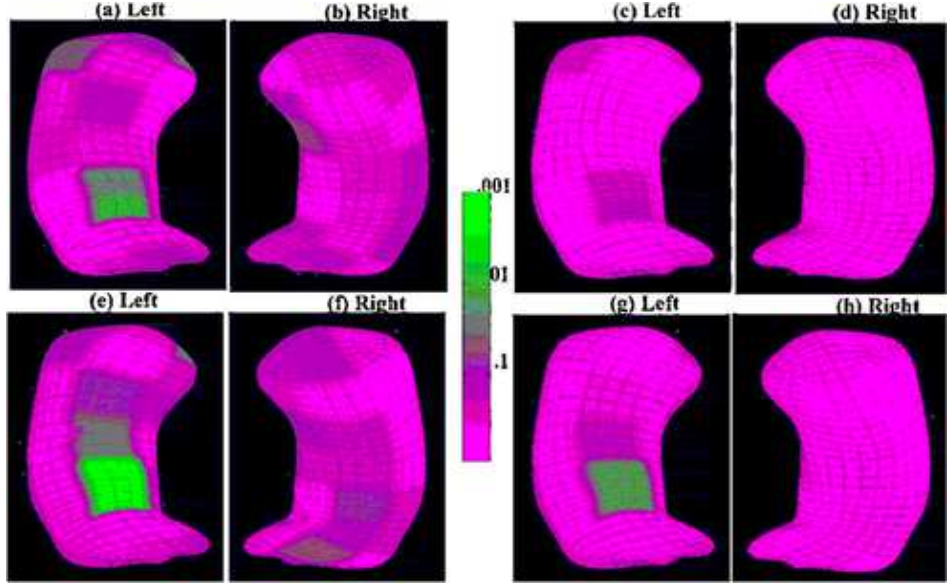}

\caption{Results from the longitudinal schizophrenia study. The top row
is for the baseline analysis: the color-coded uncorrected $p$-value
maps of group effect for \textup{(a)} the left hippocampus and \textup
{(b)} the right
hippocampus; the color-coded corrected $p$-value maps of group effect
for \textup{(c)} the left hippocampus and \textup{(d)} the right
hippocampus after
correcting for multiple comparisons. The bottom row is for the
longitudinal analysis: the color-coded uncorrected $p$-value maps of
group effect for \textup{(e)} the left hippocampus and \textup{(f)} the right
hippocampus; the color-coded corrected $p$-value maps of group effect
for \textup{(g)} the left hippocampus and \textup{(h)} the right
hippocampus after
correcting for multiple comparisons. }\label{fig6}
\end{figure}

To determine whether the changing patterns of the thickness of the
hippocampus over time are similar or not across the three groups, we
tested the null hypotheses $H_0\dvtx\beta_9=\beta_{10}=0$ ($\beta_9$ and
$\beta_{10}$ are the coefficients of the interaction terms of group
and time) at all 24 atoms for each of the left hippocampus and the
right hippocampus, and it turned out that the interaction terms were not
significant for most atoms. Next we deleted the interaction terms
and tried to look at whether there are differences in the responses
across the three groups on average over time with respect to the null
hypotheses $H_0\dvtx\beta_3=\beta_4=0$ at all 24 atoms for each of the
left hippocampus and the right hippocampus. Again we only
found that there was a significant difference through time in m-rep
thickness at the upper
central atoms in the left hippocampus across schizophrenia patients and
healthy controls groups after correcting for multiple comparisons, but the
differences were not significant at other atoms, nor at any atoms on the
right hippocampus. The color-coded $p$-values of the $\mathit
{LR}_{\mathrm{Aetel}}$
across the atoms of both the left and right reference hippocampi are
shown in Figure \ref{fig5}(e) and (f), and the corrected $p$-values
are shown in Figure \ref{fig6}(g) and (h). Before correcting for multiple
comparisons, there was a significant group difference in m-rep
thickness at the upper central atoms in the left hippocampus, and the
significance level is larger than that of the baseline analysis. Since
the positive correlation is commonly observed in imaging data, we
applied the false discovery rate (FDR) procedure in
\citet{BenYek01} to correct for multiple comparisons. There is
still a significant group effect at the upper central atoms in the left
hippocampus.

We compared the results by making the assumption that WBV was a~type
II time-dependent and also a type III time-dependent covariate.
Treating WBV as a type II time-dependent covariate lowered the $p$-values,
making some nonsignificant $p$-values for the group effect
significant. On the other hand, we found that all the
standard deviations associated with the parameter estimates
treating WBV as a type II time-dependent covariate were
uniformly less than those treating WBV as a type III,
which confirms that treating WBV as a type II gains
efficiency by making use of more correct estimating
equations. Table \ref{t:LillyStd} compares the standard deviations of
the parameter estimates between
treating WBV as a type II time-dependent covariate and a type III
time-dependent covariate at atom 11 of the left hippocampus.

%
\begin{table}
\caption{Standard deviation comparison of the parameter estimates
between treating WBV as a~type II time-dependent covariate and a~type
III time-dependent covariate at atom 11 of the left hippocampus}\label{t:LillyStd}
\begin{tabular*}{\textwidth}{@{\extracolsep{\fill}}lccccccccc@{}}
\hline
& \textbf{Intercept} & \textbf{Gender} & \textbf{Age} & \textbf{SC1} & \textbf{SC2} & \textbf{Race1} & \textbf{Race2} & \textbf{WBV} &
\textbf{Time} \\
\hline
Type III & $0.367$ & $0.078$ & $0.007$ & $0.062$ & $0.058$ &
$0.097$ & $0.102$ & $0.237$ & $0.022$\\
Type II& $0.344$ & $0.075$ & $0.005$ & $0.058$ & $0.054$ & $0.094$ &
$0.100$ & $0.221$ & $0.018$\\
\hline
\end{tabular*}
\end{table}

The longitudinal analysis increased the significance level at those
significant atoms for the group effect, compared to the baseline
analysis. We were also able to observe the change difference across
groups through time, although it is not much. Both the baseline analysis
and longitudinal analysis suggest that there is an asymmetric aspect
in that the left hippocampus shows larger regions of significance
than the right one, and the significant positions of the group
differences are around the lateral dentate gyrus and medial CA4 body
regions for the left hippocampus.

\section{Discussion}\label{sec5}

We have
developed TETEL for spatial
analysis of neuroimaging data from longitudinal
studies. We have shown that AETEL allows us to efficiently analyze
longitudinal data with different time-dependent covariate types. We have
specifically combined all the data in the closest neighborhood of each voxel
(or pixel) on a 3D volume (or 2D surface) with
appropriate weights to calculate adaptive parameter estimates and
adaptive test statistics. We have used simulation studies to examine
the finite sample performance of AETEL and TETEL.
In our longitudinal schizophrenia study, we have used the boundary and
medial shape of the hippocampus to detect differences in morphological
changes of the
hippocampus across time between schizophrenic patients and healthy
subjects. For the m-rep thickness, we have found that WBV is an
important time-dependent covariate.
Potential applications of our methodology include understanding normal
and abnormal brain development, and identifying the neural bases of the
pathophysiology and etiology of neurodegenerative and neuropsychiatric
disorders.


Many issues still merit further research. One major issue is to
develop a~test procedure based perhaps on random field theory or
resampling methods
to correct for multiple comparisons in order to control the family-wise
error rate under the moment model (\ref{ETELIMG1}).
Another major issue is to extend the test procedure to
conduct cluster size inference and examine its performance in
controlling the Type I error rate.
The test procedure may
lead to a simple cluster size test (cluster size test assesses
significance for all sizes of the connected regions greater than a
given primary threshold).
Models with nonparametric components using TETEL also may prove to be
useful directions to consider.

\begin{supplement}
\stitle{Proofs of Theorems \ref{thm1} and \ref{thm2}}
\slink[doi]{10.1214/11-AOAS480SUPP} 
\slink[url]{http://lib.stat.cmu.edu/aoas/480/supplement.pdf}
\sdatatype{.pdf}
\sdescription{We present assumptions and proofs of Theorems \ref{thm1} and \ref{thm2}.}
\end{supplement}

\section*{Acknowledgments}
We thank the Editor, an Associate Editor, and two referees for
valuable suggestions, which helped to improve our presentation greatly.
We are thankful to Sylvain
Gouttard, Steve Pizer, and Josh Levy for sharing their software.

%

\printaddresses

\end{document}